\documentclass[sigconf]{acmart}
\AtBeginDocument{%
  }

\usepackage{multirow}
\usepackage{float}
\usepackage{subfig}
\usepackage{enumitem}
\usepackage{algorithmic}
\usepackage[ruled, linesnumbered]{algorithm2e}

\author{Miaomiao Cai}
\affiliation{%
  \institution{National University of Singapore}
  \city{Singapore}
  \country{Singapore}
  }
\email{cmm.hfut@gmail.com}

\author{Zhijie Zhang}
\affiliation{%
  \institution{Hefei University of Technology}
  \city{Hefei}
  \country{China}}
\email{zhijiezhang021@gmail.com}

\author{Junfeng Fang}
\affiliation{%
  \institution{National University of Singapore}
  \city{Singapore}
  \country{Singapore}}
\email{fangjf1997@gmail.com}

\author{Zhiyong Cheng}
\affiliation{%
  \institution{Hefei University of Technology}
  \city{Hefei}
  \country{China}}
\email{jason.zy.cheng@gmail.com}

\author{Xiang Wang}
\affiliation{%
  \institution{University of Science and Technology of China	}
  \city{Hefei}
  \country{China}}
\email{xiangwang1223@gmail.com}

\author{Meng Wang}
\authornote{Corresponding Authors.}
\affiliation{%
  \institution{Hefei University of Technology}
  \city{Hefei}
  \country{China}}
\email{eric.mengwang@gmail.com}

\copyrightyear{2026}
\acmYear{2026}
\setcopyright{cc}
\setcctype{by}
\acmConference[WWW '26]{Proceedings of the ACM Web Conference 2026}{April 13--17, 2026}{Dubai, United Arab Emirates}
\acmBooktitle{Proceedings of the ACM Web Conference 2026 (WWW '26), April 13--17, 2026, Dubai, United Arab Emirates}
\acmPrice{}
\acmDOI{10.1145/3774904.3792617}
\acmISBN{979-8-4007-2307-0/2026/04}

\renewcommand{\shortauthors}{Miaomiao et al.}

\begin{document}

\title{RMBRec: Robust Multi-Behavior Recommendation towards Target Behaviors}

\newcommand{\fullname}{\textit{\textbf{R}obust \textbf{M}ulti-\textbf{B}e\-hav\-ior \textbf{Rec}\-om\-men\-da\-tion towards Target Behaviors (RMBRec)}}
\newcommand{\shortname}{\emph{RMBRec}}

\begin{abstract}

Multi-behavior recommendation faces a critical challenge in practice: auxiliary behaviors (e.g., clicks, carts) are often noisy, weakly correlated, or semantically misaligned with the target behavior (e.g., purchase), which leads to biased preference learning and suboptimal performance. While existing methods attempt to fuse these heterogeneous signals, they inherently lack a principled mechanism to ensure robustness against such behavioral inconsistency.

In this work, we propose ~\fullname, a robust multi-behavior recommendation framework grounded in an information-theoretic robustness principle. We interpret robustness as a joint process of maximizing predictive information while minimizing its variance across heterogeneous behavioral environments. Under this perspective, the \textbf{Representation Robustness Module (RRM)} enhances \textit{local semantic consistency} by maximizing the mutual information between users' auxiliary and target representations, whereas the \textbf{Optimization Robustness Module (ORM)} enforces \textit{global stability} by minimizing the variance of predictive risks across behaviors, which is an efficient approximation to invariant risk minimization.
This local–global collaboration bridges representation purification and optimization invariance in a theoretically coherent way. Extensive experiments on three real-world datasets demonstrate that RMBRec not only outperforms state-of-the-art methods in accuracy but also maintains remarkable stability under various noise perturbations.
For reproducibility, our code is available at \url{https://github.com/miaomiao-cai2/RMBRec/}. 


\end{abstract}



\maketitle

\section{Introduction}

Recommender systems (RS) have become essential tools of the Web ecosystem~\cite{DBLP:journals/tois/ZhuoQHDLW24,xu2025fair,cheng2018aspect}, helping users discover preferred items, news, or services from massive candidate pools~\cite{chen2019efficient,covington2016deep}. Traditional studies usually focus on a single type of feedback (e.g., clicks or purchases), a setting often referred to as \textit{single-behavior recommendation}~\cite{Rendle2009BPRBP,Chen2020RevisitingGB,Koren2009MatrixFT}. However, user interactions in real-world scenarios are inherently \textit{multi-dimensional}, including view, collect, add-to-cart, and purchase behaviors~\cite{jin2020multi,he2024denoising,ma2025graph,han2024efficient}. Leveraging such heterogeneous feedback, known as \textit{multi-behavior recommendation}, can enrich supervision and improve the prediction of target behaviors (e.g., purchases)~\cite{wu2024multi}. 

\begin{figure}
    \centering
    \vspace{-0.2cm}
    \includegraphics[width=\linewidth]{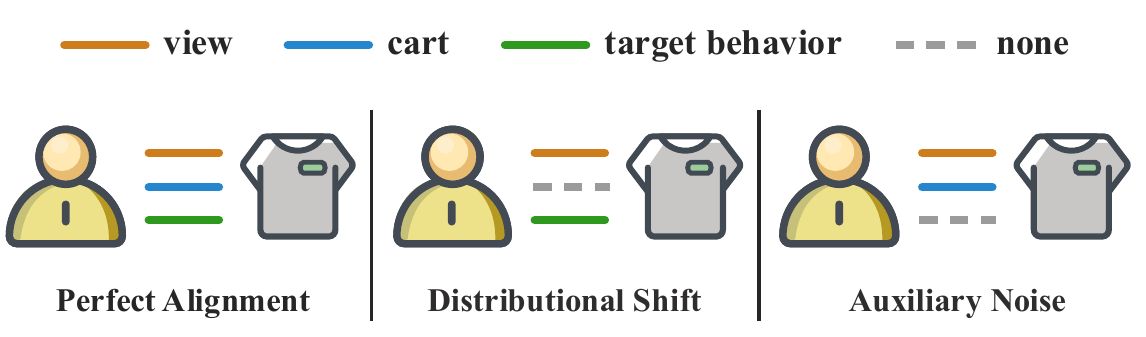}
    \caption{Illustration of distributional shift and auxiliary noise in multi-behavior recommendation.}
    \label{fig:intro}
    \vspace{-0.8cm}
\end{figure}

Despite this potential, most existing multi-behavior recommendation works rely on fusion-based paradigm, which learns behavior-specific embeddings and aggregates them through concatenation~\cite{Chen2020RevisitingGB,He2020LightGCNSA}, MLPs~\cite{han2024efficient,gao2019learning}, or attention mechanisms~\cite{jin2020multi,meng2023parallel,ma2025graph}. In practice, however, auxiliary interactions (e.g., view or cart) are often noisy or weakly correlated with the target behavior~\cite{han2024efficient,cai2025neighborhood,wang2025unleashing,zhao2025symmetric}. Users may casually browse or add items to carts without genuine purchase intent, introducing inconsistent supervision signals that distort the learned preference space. As a result, multi-behavior recommenders tend to overfit spurious correlations rather than capture true purchasing intent, leading to poor robustness in real-world environments~\cite{wang2024distributionally,o2004collaborative}. This vulnerability mainly arises from two types of inconsistencies, as shown in Figure~\ref{fig:intro}. (1) \textbf{Distributional shift:} some users directly purchase without performing auxiliary actions, causing the distributions of auxiliary and target behaviors to diverge~\cite{yan2025user,cai2025neighborhood}. (2) \textbf{Auxiliary noise:} many auxiliary interactions (e.g., random views or abandoned carts) are unrelated to real purchase intent~\cite{zhang2023denoising,zhao2025symmetric}. These two factors jointly create semantic misalignment across behaviors, making naive fusion of heterogeneous signals unreliable.

To quantify the severity of such misalignment, we introduce a diagnostic metric called the \textbf{Behavioral Alignment Ratio (BAR)}, which measures the degree to which each auxiliary behavior aligns with the target one. BAR reveals significant disparities across datasets, indicating that auxiliary signals differ in reliability. For example, in some datasets nearly all purchases follow auxiliary actions (\textit{high BAR}), while in others the majority of purchases occur independently (\textit{low BAR}). This observation highlights an insight that a robust model must exploit well-aligned signals while resisting noisy or misaligned ones. From an information-theoretic perspective~\cite{Wu2020SelfsupervisedGL,yang2021enhanced}, robustness in multi-behavior learning requires maximizing the predictive information from reliable behaviors while minimizing the influence of noisy or distributionally shifted ones.

To achieve this goal, we propose~\fullname, whose core motivation is that auxiliary behaviors differ widely in both semantic relevance and noise level. Guided by this principle, we design two collaborative modules that jointly enhance robustness from local semantic consistency to global optimization stability, grounded in an information-theoretic robustness formulation~\cite{Wu2020SelfsupervisedGL,yang2021enhanced}.

\textbf{(1) Representation Robustness Module (RRM).} 
Instead of treating all behavior-specific embeddings equally, we anchor the entire representation learning process on the target behavior. Specifically, we propose a target-oriented contrastive learning objective that operates solely on user embeddings, based on the key insight that semantic drift primarily occurs in user representations. 
While items maintain relatively stable characteristics across different behaviors (an item's properties remain consistent whether it is viewed, carted, or purchased), user intent and preference expression vary significantly across behavior types. A user may casually view or add items to cart without genuine purchase intent, creating semantic inconsistency in their representation across behaviors. 
Our contrastive objective pulls the user embeddings from auxiliary behaviors (e.g., view, cart) closer to their counterpart from the target behavior (purchase), while pushing away embeddings from irrelevant users. This process can be viewed as \textit{maximizing the mutual information} between auxiliary and target representations, thereby enforcing \textbf{local semantic consistency} and purifying noisy supervision signals where misalignment is most severe. 

\textbf{(2) Optimization Robustness Module (ORM).} 
We further view each behavior type (view, cart, etc.) as a distinct behavioral environment with its own distribution and noise pattern. Drawing inspiration from invariant learning~\cite{arjovsky2019invariant,yang2025invariance,liao2025mitigating}, we introduce a regularization term that enforces the model to learn invariant user preferences across these environments. 
From an information-theoretic perspective~\cite{Wu2020SelfsupervisedGL,yang2021enhanced}, ORM aims to \textit{minimize the variance of predictive information} across behaviors, ensuring that the captured patterns correspond to stable and causal preference factors rather than behavior-specific shortcuts. Through invariance regularization, ORM provides \textbf{global optimization stability} and prevents the model from overfitting to any particular behavioral distribution.

To this end, RRM and ORM form a unified robustness framework that bridges \textit{local semantic consistency} and \textit{global optimization stability}. They jointly implement the principle of \textbf{maximizing predictive information while minimizing its variance} across heterogeneous behaviors, offering a principled solution to behavioral noise and misalignment in multi-behavior recommendation.
We conduct extensive experiments on three real-world datasets. The results demonstrate that ~\shortname ~ significantly outperforms state-of-the-art baselines. More importantly, extensive ablation and noise-injection experiments are conducted to validate the robustness of our model against noisy environments.
The main contributions are summarized as follows: 

\begin{itemize}[leftmargin=0.3cm, itemindent=0cm]

    \item  We propose a target-anchored contrastive learning mechanism to purify auxiliary behaviors and enhance local semantic consistency on the user side with the target behavior in the representation space.
    
    \item  We introduce an invariant optimization regularizer that treats different behaviors as heterogeneous environments, thereby ensuring global optimization stability and enhanced robustness against noisy supervision.
    
    \item We demonstrate that ~\shortname~ achieves superior accuracy and robustness over state-of-the-art baselines on three datasets. Its effectiveness is further validated through in-depth ablation studies and noise-injection tests.
    
\end{itemize}

\section{Related Works}

\textbf{Multi-Behavior Recommendation.} Multi-behavior recommendation (MBR) models user preferences by leveraging heterogeneous interactions such as \textit{view}, \textit{cart}, and \textit{purchase}~\cite{gao2019learning, jin2020multi}. Early methods extend matrix factorization or collaborative filtering by sharing latent factors across behaviors~\cite{gao2019learning}. With the advancement of graph neural networks (GNNs), subsequent studies construct behavior-specific graphs and perform cross-behavior propagation to capture high-order user–item dependencies~\cite{jin2020multi, cheng2023multi, yan2023cascading,liu2021interest}. Attention-based approaches further adaptively weigh different behaviors to enhance flexibility~\cite{yan2024behavior, meng2023parallel}. 
Despite their effectiveness, most existing MBR methods implicitly assume that auxiliary behaviors are semantically consistent with the target behavior. In practice, however, auxiliary interactions are often noisy or weakly correlated with target preferences (e.g., frequent views without purchases), which may introduce biased supervision and degrade robustness~\cite{gao2022self,li2023multiple,he2024double,Liu2016MRELBP,LiSARATRX25}. 

\textbf{Contrastive Learning for Multi-Behavior Recommendation.}
To enhance robustness against noisy or insufficient supervision, recent recommendation research has widely adopted contrastive learning to regularize latent representations through auxiliary consistency signals~\cite{chuang2022robust,wei2021contrastive,cai2024mitigating,cai2024popularity,cai2025graph}. In single-behavior settings, SimGCL~\cite{Yu2021AreGA} improves robustness and generalization by perturbing user and item embeddings with stochastic noise and enforcing InfoNCE-based contrastive consistency. 
Building on this paradigm, contrastive learning has been extended to multi-behavior recommendation, where heterogeneous behaviors are modeled via graph propagation or shared latent structures. Representative methods such as CRGCN~\cite{yan2023cascading} and MB-CGCN~\cite{cheng2023multi} refine cross-behavior dependencies through cascading or residual propagation, while PKEF~\cite{meng2023parallel} adopts a parallel expert fusion framework to disentangle and balance heterogeneous behavior signals. Furthermore, self-supervised approaches including S-MBRec~\cite{gu2022self}, MBSSL~\cite{xu2023multi}, and MISSL~\cite{wu2024multi} introduce intra- and inter-behavior contrastive objectives to enhance semantic consistency and optimization stability under noisy supervision. Meanwhile, BCIPM~\cite{yan2024behavior} incorporates behavior-contextualized preference modeling to filter task-irrelevant signals. 
Despite these advances, most existing methods still regularize representations via contrastive or reconstruction objectives, implicitly assuming consistent cross-behavior semantics, and thus remain limited in exploiting intrinsic behavioral heterogeneity to address behavior-induced optimization instability and achieve robust generalization.

\textbf{Invariant Learning for Multi-Behavior Recommendation.}
Invariant representation learning~\cite{arjovsky2019invariant, krueger2021out} aims to identify causal factors that remain stable across different environments, thereby improving robustness under distributional shifts. This principle has recently been explored in recommendations to mitigate bias and domain shift~\cite{yang2025invariance, liao2025mitigating,zhang2023invariant}. UIPL~\cite{yan2025user} follows the invariant risk minimization (IRM) framework to learn user-invariant preferences from multi-behavior data. It constructs synthetic environments by combining multiple behaviors and employs a variational autoencoder (VAE) ~\cite{kingma2013auto}to generate latent invariant representations optimized. However, the environments in UIPL are artificially constructed rather than derived from the naturally heterogeneous behavioral environments inherent in multi-behavior recommendation. Moreover, its generative framework introduces considerable modeling complexity and constrains invariance to the representation space, without addressing the optimization instability across behaviors. 

In summary, while prior studies have enhanced robustness through denoising, contrastive, or representation-level invariance, they typically assume consistent cross-behavior semantics and overlook the potential of leveraging natural behavioral heterogeneity to achieve optimization-level stability and robust generalization across diverse behaviors in real-world settings.

\begin{figure*}[t]
    \centering
    \vspace{-0.5cm}
    \includegraphics[width=0.9\linewidth]{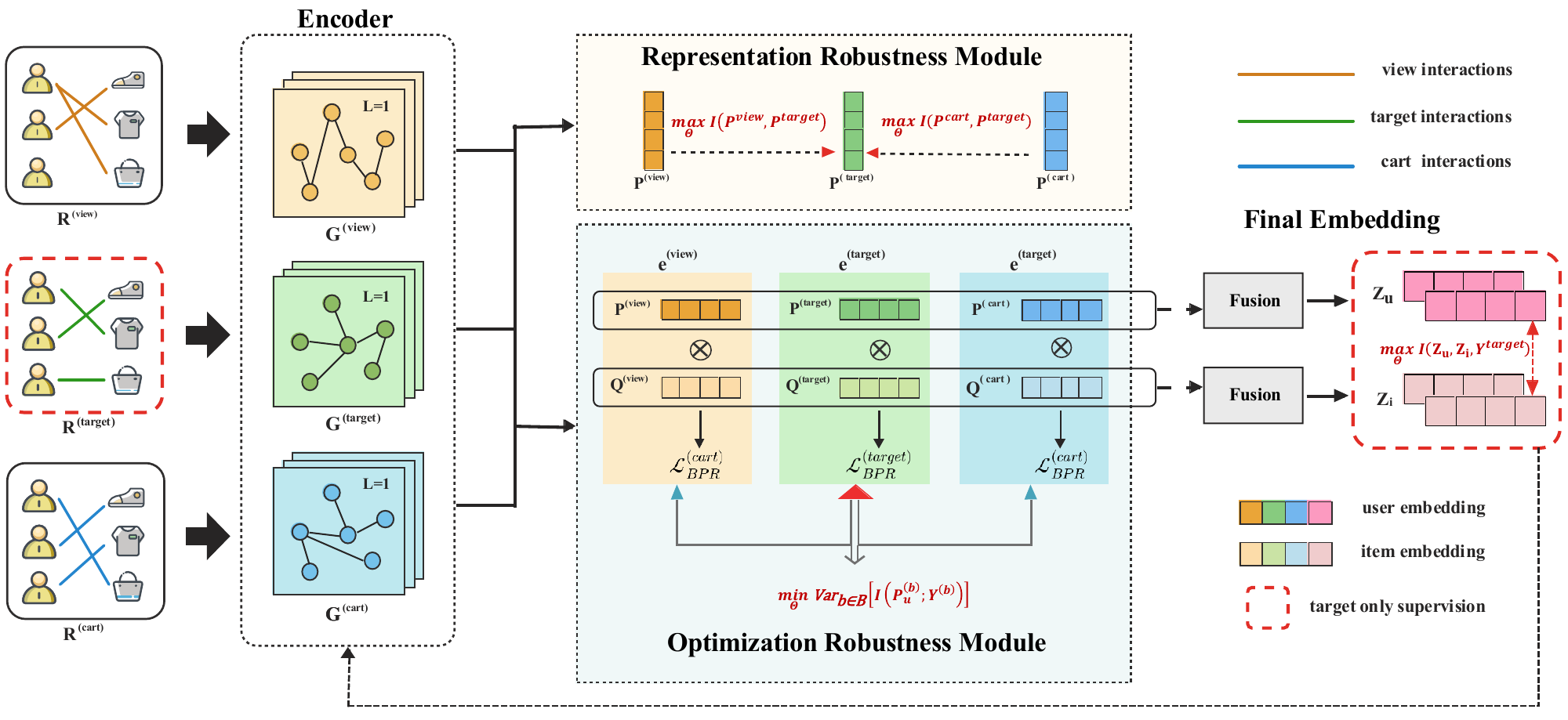}
    \caption{The overview of~\shortname, which improves robustness via Representation Robustness and Optimization Robustness.}
    \label{fig:framework}
    \vspace{-0.4cm}
\end{figure*}

\section{Model}

In this section, we present the proposed~\shortname~framework. As illustrated in Figure~\ref{fig:framework}, our model consists of two complementary modules operating at different levels to enhance robustness: (1) the \textbf{Representation Robustness Module (RRM)}, which ensures local semantic consistency by maximizing the mutual information between auxiliary and target representations; and (2) the \textbf{Optimization Robustness Module (ORM)}, which enforces global optimization stability by minimizing the variance of predictive risks across behavioral environments. Together, they instantiate the information-theoretic~\cite{Wu2020SelfsupervisedGL,yang2021enhanced} robustness principle of \textit{maximizing predictive information while minimizing its variance} across heterogeneous behaviors.

\subsection{Multi-Behavior Encoding}

Let $\mathcal{U}$ and $\mathcal{I}$ denote the sets of users and items, respectively. Each user $u \in \mathcal{U}$ may interact with item $i \in \mathcal{I}$ through multiple types of behaviors $\mathcal{B} = \{view, collect, add-to-cart, purchase, \ldots\}$, with purchase as the target behavior. For each behavior $b \in \mathcal{B}$, the interaction matrix is denoted as $\mathbf{R}^{(b)} \in \{0,1\}^{|\mathcal{U}| \times |\mathcal{I}|}$.
Our goal is to learn unified user and item embeddings $\mathbf{z}_u$ and $\mathbf{z}_i$ that effectively predict target behavior interactions. We begin by initializing the user embedding matrix $\mathbf{Z}_u \in \mathbb{R}^{|\mathcal{U}| \times d}$ and item embedding matrix $\mathbf{Z}_i \in \mathbb{R}^{|\mathcal{I}| \times d}$, which serves as the foundation for multi-behavior representation learning. 

For each behavior type $b \in \mathcal{B}$, we employ an encoder $f(\cdot)$ to learn behavior-specific representations. 
The encoding process generates behavior-specific user and item embedding matrices from the interaction graph and unified embeddings: 

\begin{equation}
    \mathbf{P}^{(b)},\mathbf{Q}^{(b)} = f(\mathbf{R}^{(b)}, \mathbf{Z}_u, \mathbf{Z}_i),
\end{equation}
where $\mathbf{P}^{(b)} \in \mathbb{R}^{|\mathcal{U}| \times d}$ and $\mathbf{Q}^{(b)} \in \mathbb{R}^{|\mathcal{I}| \times d}$ are the behavior-specific user and item embedding matrices for behavior $b$, respectively. In our implementation, we choose LightGCN as the encoder due to its effectiveness and efficiency in capturing high-order collaborative signals through simplified graph propagation~\cite{Wang2019NeuralGC,He2020LightGCNSA}.

For each behavior $b$, we compute a behavior-specific BPR loss~\cite{Rendle2009BPRBP} to capture the preference relationships within that behavior:

\begin{equation}
    \mathcal{L}^{(b)}_{BPR} = - \frac{1}{|\mathcal{O}^{(b)+}|} \sum_{(u,i,j) \in \mathcal{O}^{(b)+}} \log \sigma\left({\mathbf{p}_u^{(b)}}^\top \mathbf{q}_i^{(b)} - {\mathbf{p}_u^{(b)}}^\top \mathbf{q}_j^{(b)}\right),
\end{equation}
where $\mathcal{O}^{(b)+} = \{(u,i,j) \mid \mathbf{R}^{(b)}_{u,i}=1, \mathbf{R}^{(b)}_{u,j}=0\}$ contains training triplets for behavior $b$, $\mathbf{p}_u^{(b)}$ denotes the row vector of user $u$ in $\mathbf{P}^{(b)}$, and $\sigma(\cdot)$ is the sigmoid function.

\subsection{Representation Robustness Module (RRM)}

In multi-behavior recommendation, semantic drift is inherently user-centric. While item properties remain invariant across behaviors (e.g., an item's intrinsic attributes stay consistent whether it is merely viewed or purchased), user intent varies significantly. For example, users may casually browse or add items to carts without genuine purchase intention, which induces stochastic noise and semantic drift specifically in user representations. To address this user-side misalignment, we introduce RRM to enhance local semantic consistency. It maximizes the predictive information between auxiliary and target user representations, i.e., $I(P_u^{(b)}; P_u^{(t)})$, ensuring that the learned embeddings retain target-relevant signals while suppressing behavior-specific noise. In particular, RRM realizes this objective via a target-anchored contrastive formulation.

Inspired by the success of contrastive learning in alignment representation learning~\cite{khosla2020supervised,wang2020understanding,yang2023generative,Wu2020SelfsupervisedGL}, we design a target-anchored contrastive alignment objective to solely guide auxiliary user embeddings toward the target semantics.
Specifically, for each user $u$, the target-behavior embedding $\mathbf{p}_u^{(target)}$ serves as a reliable anchor that reflects genuine intent. For any auxiliary behavior $b \in \mathcal{B}_{aux}$, we construct a positive pair $(\mathbf{p}_u^{(b)}, \mathbf{p}_u^{(target)})$ to enforce semantic alignment between auxiliary and target representations. The negative samples are embeddings of other users under the same auxiliary behavior $b$ within the same mini-batch, i.e., $(\mathbf{p}_u^{(b)}, \mathbf{p}_{u'}^{(b)})$ with $u' \neq u$. This in-batch sampling strategy avoids the need for a global memory bank and focuses optimization on user representations only, which is both semantically meaningful and computationally efficient. By aligning user embeddings locally within each auxiliary behavior, RRM captures user-level semantic consistency while maintaining low training overhead. Formally, the contrastive loss for each auxiliary behavior $b$ of a user $u$ is defined as: 

\begin{equation}
\label{eq:rrm_loss_unit}
\mathcal{L}_{RRM}^{(b,u)} = - \log \frac{\exp(sim(\mathbf{p}_u^{(b)}, \mathbf{p}_u^{(target)}) / \tau)}{\sum_{u' \in \mathcal{N}_u} \exp(sim(\mathbf{p}_u^{(b)}, \mathbf{p}_{u'}^{(b)}) / \tau)},
\end{equation}
where \( sim(\cdot, \cdot) \) is the cosine similarity function, \( \tau \) is a temperature hyperparameter that controls the sharpness of the distribution, and \( \mathcal{N}_u \) is the set of in-batch negative users. 
The overall RRM loss is then averaged over all users and all auxiliary behaviors:
\begin{equation}
\label{eq:rrm_loss}
\mathcal{L}_{RRM} =  \frac{1}{|\mathcal{B}|-1} \frac{1}{\mathcal{U}}\sum_{b\in \mathcal{B}\setminus \{target\}}\sum_{u\in \mathcal{U}}\mathcal{L}_{RRM}^{(b, u)}.
\end{equation}
This loss explicitly aligns auxiliary behavior embeddings toward the target semantics, ensuring that informative signals are strengthened while semantically drifted ones are suppressed. 
By maximizing the expected mutual information between auxiliary and target embeddings, RRM enforces user-level semantic alignment, filtering noisy auxiliary signals and establishing \textbf{representation-level (local) robustness} under distribution shifts.


\subsection{Optimization Robustness Module (ORM)}

While RRM addresses semantic noise, distributional shift across behavioral domains may still degrade generalization. Different behaviors (e.g., view, cart, purchase) follow distinct distributions and noise patterns, introducing environment-specific biases.  For example, views are often biased by exposure, carts reflect tentative interests, and purchases represent genuine intent. 
To achieve robustness beyond local semantic consistency, we introduce the \textbf{Optimization Robustness Module (ORM)}.
It minimizes the variance of predictive information $I(P_u^{(b)};Y^{(b)})$ across behaviors, encouraging the model to depend on causal preference features rather than behavior-specific correlations. Specifically, it realizes this goal through an invariant learning framework that regularizes optimization across heterogeneous behaviors.

\subsubsection{Behavioral Environments as Optimization Domains}

Drawing inspiration from Invariant Risk Minimization (IRM)~\cite{arjovsky2019invariant,yang2025invariance,liao2025mitigating}, we refine multi-behavior recommendation as a multi-environment learning problem, where each behavior type defines a distinct \textit{behavioral environment}:

\begin{equation}
    \mathcal{E} = \{ e^{(b)} \mid b \in \mathcal{B} \}, \quad e^{(b)} \sim P(\mathbf{R}^{(b)}).
\end{equation}
Each behavioral environment $e^{(b)}$ corresponds to one behavior type (e.g., view, cart, purchase), and the corresponding interaction graph $\mathbf{R}^{(b)}$ defines its data distribution. For example, the \textit{view environment} is dominated by high-volume, low-commitment interactions with strong popularity bias, while the \textit{purchase environment} contains sparse but reliable signals of the users' true preference.  
Although the marginal distributions $P(\mathbf{R}^{(b)})$ differ across environments, the underlying user preference mechanism should ideally remain invariant across behavior types.

To capture these invariant preferences, ORM introduces a shared encoder $h_\phi$ and a shared predictor $w$. For each behavioral environment $e^{(b)}$, the empirical risk is $\mathcal{L}^{(b)}_{BPR}(w \circ h_\phi)$. The optimization objective follows the IRM principle~\cite{arjovsky2019invariant,yang2025invariance}: 

\begin{equation} 
    \min_{\phi,\, w} \sum_{e^{(b)} \in \mathcal{E}} \mathcal{L}^{(b)}_{BPR}(w \circ h_{\phi}) \quad \text{s.t. } \nabla_{w} \mathcal{L}^{(b)}_{BPR}(w \circ h_{\phi}) = \mathbf{0}. 
    \label{eq:irm} 
\end{equation} 

This constraint enforces that the same predictor $w$ remains optimal across all behavioral environments. If a single $w$ performs consistently well in diverse environments, $h_\phi$ must have captured the invariant and causal preference features rather than behavior-specific noise. 

\subsubsection{Learning Invariant Preferences for Global Robustness}

While the ideal IRM objective in Equation~\ref{eq:irm} provides a principled formulation, it is computationally intractable to directly optimize this objective~\cite{arjovsky2019invariant,yang2025invariance,liao2025mitigating}. 
Following prior work~\cite{yang2025invariance,liao2025mitigating,chen20253}, we adopt the Risk Extrapolation (REx) approximation~\cite{krueger2021out}, which empirically achieves more stable and efficient optimization than IRMv1~\cite{arjovsky2019invariant} or IRMv2~\cite{ahuja2020invariant}, while maintaining the same goal of minimizing performance disparity across environments.
Following the IRM principle~\cite{arjovsky2019invariant}, our objective is to learn a preference prediction function that performs consistently well across all behavioral environments. 
This ensures that the model depends only on features, maintaining stable relationships with the target behavior, rather than exploiting behavior-specific correlations.

ORM enforces invariance by penalizing the variance of these risks/loss across behavioral environments $\mathcal{L}^{(b)}_{BPR}$: 
\begin{equation}
\label{eq:orm_loss}
\mathcal{L}_{ORM} = Var\left(\{\mathcal{L}^{(b)}_{BPR}\}_{b \in \mathcal{B}}\right) 
= \frac{1}{|\mathcal{B}|} \sum_{b \in \mathcal{B}} \left(\mathcal{L}^{(b)}_{BPR} - \bar{\mathcal{L}}\right)^2,
\end{equation}
where $\bar{\mathcal{L}} = \frac{1}{|\mathcal{B}|} \sum_{b \in \mathcal{B}} \mathcal{L}^{(b)}_{BPR}$ is the mean loss across all behaviors.

Minimizing $\mathcal{L}_{ORM}$ encourages the model to achieve uniform predictive performance across all environments, thus preventing overfitting to behavior-specific spurious patterns (e.g., “frequent viewers but non-buyers”). This process explicitly guides the optimization toward stable solutions that generalize beyond environment-specific distributions and spurious correlations. 

ORM enhances robustness at the optimization level by stabilizing the training across heterogeneous behaviors. Unlike prior methods, ORM does not rely on explicit noise labels or handcrafted priors. Instead, it leverages the natural heterogeneity of user behaviors as a supervised signal for invariance learning. In combination with RRM which purifies embeddings through target-anchored contrastive alignment, In combination with RRM which purifies embeddings through target-anchored contrastive alignment,~\shortname ~ ensures that the model maintains both \textbf{local semantic consistency} and \textbf{global optimization stability}, achieving robustness from representation to optimization in a unified and principled manner. 

\subsection{Target Prediction and Optimization}

After obtaining the purified behavior-specific embeddings through our RRM and ORM modules, we fuse them to obtain the final unified representations. We employ a simple yet effective fusion~\cite{He2020LightGCNSA}:
\begin{equation}
\mathbf{z}_u = \frac{1}{|\mathcal{B}|} \sum_{b \in \mathcal{B}} \mathbf{p}_u^{(b)}, \quad
\mathbf{z}_i = \frac{1}{|\mathcal{B}|} \sum_{b \in \mathcal{B}} \mathbf{q}_i^{(b)}.
\end{equation}

We employ equal-weight fusion to aggregate behavior-specific embedding. This equal-weight fusion ensures that all behavior types contribute equally to the final representation, while the purification through RRM and ORM ensures that only semantically aligned and optimizational stable signals are preserved. Using these fused embeddings, we compute the final preference score for target behavior prediction: $s(u,i) = \mathbf{z}_u^\top \mathbf{z}_i$.

The main optimization objective follows the Bayesian Personalized Ranking (BPR) framework~\cite{Rendle2009BPRBP}, using only the target behavior interactions as supervision:

\begin{equation} 
    \mathcal{L}_{main} = - \frac{1}{|\mathcal{O}^+|} \sum_{(u,i,j) \in \mathcal{O}^+} \log \sigma\left(s(u,i) - s(u,j)\right)+\lambda ||\Theta||^2_2,
    \label{equa:bpr_main}
\end{equation}
where $\mathcal{O}^+ = \{(u,i,j) \mid \mathbf{R}^{(target)}_{u,i}=1, \mathbf{R}^{(target)}_{u,j}=0\}$ contains training triplets constructed from target behavior interactions, and $\Theta$ denotes all trainable parameters.

The complete training objective integrates the main recommendation loss with our two proposed robustness modules. This unified formulation combines: (1) the primary target behavior supervision $\mathcal{L}_{main}$ anchors optimization to the target task and prevents representation drift, (2) the representation robustness loss $\mathcal{L}_{RRM}$ enforces local semantic consistency through target-anchored contrastive learning, and (3) the optimization robustness loss $\mathcal{L}_{ORM}$ ensures global stability through invariance regularization across behavioral environments.

\begin{equation}
    \mathcal{L} = \mathcal{L}_{main} + \lambda_1 \mathcal{L}_{RRM} + \lambda_2 \mathcal{L}_{ORM},
\end{equation}
where hyperparameters $\lambda_1$ and $\lambda_2$ balance the contributions of the robustness modules.

\subsection{Information-Theoretic Analysis}
From an information-theoretic perspective~\cite{Wu2020SelfsupervisedGL,yang2021enhanced}, robust multi-behavior learning can be viewed as jointly maximizing and stabilizing the predictive information contained in user embeddings.
Specifically, the \textbf{RRM} module maximizes the mutual information between auxiliary and target representations, i.e., $I(P_u^{(b)}; P_u^{(t)})$, which purifies auxiliary signals by enhancing local semantic consistency.
In contrast, the \textbf{ORM} module regularizes the variance of the predictive information $I(P_u^{(b)}; Y^{(b)})$ across heterogeneous behaviors, enforcing global stability of the learned representations.
These two modules form a collaborative robustness principle summarized as:

\[
\max_{\Theta}\; \Big( \underbrace{\mathbb{E}_{b\in\mathcal{B}_{\text{aux}}} [I(P_u^{(b)};P_u^{(t)})]}_{\text{(1) local MI alignment}} - \lambda\,\underbrace{\mathrm{Var}_{b\in\mathcal{B}} [I(P_u^{(b)};Y^{(b)})]}_{\text{(2) global MI invariance}} \Big). 
\] 

where the first term ensures \textit{semantic alignment} between behaviors, and the second term guarantees \textit{environmental invariance} of predictive information. This unified view bridges representation-level purification and optimization-level stability, providing a theoretical rationale for the robustness of ~\shortname. And in practice, we formally interpret RRM and ORM as optimizing the expectation and variance components of predictive mutual information. A detailed theoretical analysis, algorithm, and additional discussion on model limitations are provided in Appendix~\ref{appendix:model}.  

\section{Experiments}

In this section, we evaluate the effectiveness and robustness of the proposed model~\shortname. We first introduce the datasets, baselines, and evaluation protocols. Then, we present overall performance comparisons with state-of-the-art methods, followed by ablation studies that analyze the effects of key components and auxiliary behaviors. We further examine the impact of invariance regularization, test robustness under different noise, and conduct hyperparameter sensitivity analysis.

\subsection{Experiments Settings}

\subsubsection{Datasets}

\begin{table*}[t]
    \centering
    \vspace{-0.3cm}
    \caption{Statistics of three datasets ("--" means the dataset does not contain this behavior).}
    \resizebox{\linewidth}{!}{
    \label{tab:dataset}
        \begin{tabular}{lcccccccc}
        \toprule
        \textbf{Dataset} & \#\textbf{User} & \#\textbf{Item} & \#\textbf{Interaction} & \textbf{Behavior Type} &\textbf{$\textbf{BAR(View)}$}& \textbf{$\textbf{BAR(Collect)}$} & \textbf{$\textbf{BAR(Add-to-cart)}$}  &\textbf{$\textbf{DT}$} \\
        \midrule
        \textbf{Taobao}  & 48,749  & 39,493  &2,001,680  &$\{view,add-to-cart,purchase\}$&0.6349  &   --   &0.1509  &0.3259  \\
        \textbf{TMall}   & 41,738  & 11,953  &2,324,166 &$\{view,add-to-cart,collect,purchase\}$ &0.8213  &0.1009  &0.0012  &0.1617  \\
        \textbf{Beibei}  & 21,716  & 7,977   &3,359,784 &$\{view,add-to-cart,purchase\}$ &0.9756  &   --   &0.9731  &0.0244  \\
        \bottomrule
        \end{tabular}}
    \vspace{-0.3cm}
\end{table*}

We evaluate our model ~\shortname ~ on three widely used multi-behavior datasets: \textbf{Taobao\footnote{https://tianchi.aliyun.com/dataset/649}}, \textbf{TMall\footnote{https://tianchi.aliyun.com/dataset/140281}}, and \textbf{Beibei}. Table~\ref{tab:dataset} summarizes their statistics, including the number of users, items, interactions, and behavior type. In all datasets, \textbf{purchase} is regarded as the target behavior, while the others (e.g., view, add-to-cart, collect) are treated as auxiliary behaviors. 

To quantitatively assess the consistency between auxiliary and target behaviors, we propose the \textbf{Behavioral Alignment Ratio (BAR)} to guide robustness analysis as:
\begin{equation}
    BAR(b) = P(b|target)=\frac{|\mathbf{R}^{(b)} \cap \mathbf{R}^{(target)}|}{|\mathbf{R}^{(target)}|}, b \in \mathcal{B},
    \label{eq:bar}
\end{equation}
where $\mathbf{R}^{(b)}$ denotes the set of user–item interactions under auxiliary behavior $b$, and $\mathbf{R}^{(target)}$ represents the set of target interactions. 
Intuitively, $BAR(b)$ captures the proportion of target interactions that are accompanied by a given auxiliary behavior. 
Unlike conventional conversion rates (which correspond to $P(target\,|\,auxiliary)$), $BAR$ reflects the inverse perspective $P(auxiliary\,|\,target)$, emphasizing \textit{\textbf{how much of the target behavior can be explained by a certain auxiliary signal}}. This target-oriented formulation is particularly suitable for robustness analysis. In multi-behavior recommendation, the objective is to predict the target behavior (e.g., purchase). Thus, it is more meaningful to quantify how reliably each auxiliary behavior aligns with the target rather than how often auxiliary actions lead to purchases. A high $BAR(b)$ indicates that auxiliary behavior $b$ consistently co-occurs with the target, implying strong semantic relevance and low noise; conversely, a low $BAR(b)$ reveals weak or misleading behavioral alignment. 

To complement this measure, we also report the proportion of Direct Target (DT) behaviors—i.e., target interactions without any preceding auxiliary behavior. This ratio reflects how frequently purchases occur independently of other signals, thereby indicating the level of randomness or decision noise in user actions. 

As shown in Table~\ref{tab:dataset}, behavioral alignment exhibits striking heterogeneity across datasets. In Beibei, both view and add-to-cart behaviors display extremely high $BAR$ values ($0.9756$ and $0.9731$), while $DT$ is negligible ($0.0244$). This suggests that purchases are almost always accompanied by consistent preparatory behaviors, implying a highly coherent and low-noise environment. In contrast, Taobao and TMall exhibit significantly lower $BAR$ values, particularly for add-to-cart in TMall ($0.0012$), indicating that auxiliary behaviors are often weakly correlated or even semantically irrelevant to the target. Their relatively large $DT$ ratios further reveal that users frequently purchase directly without prior interactions, \textit{\textbf{leading to unstable supervision and distributional inconsistency}}in real-world settings.

\subsubsection{Baselines and Implementation Details}

We compare ~\shortname~\\ with a comprehensive set of representative baselines from two main categories: (1) \textbf{\textit{single-behavior methods}} and (2) \textbf{\textit{multi-behavior methods}}. The single-behavior models ~(LightGCN~\cite{He2020LightGCNSA} and SimGCL\\~\cite{Yu2021AreGA}  ~) serve as strong baselines for modeling target-only interactions, where SimGCL introduces contrastive regularization to enhance representation robustness. Among multi-behavior models, earlier graph-based methods such as CRGCN~\cite{yan2023cascading}, MB-CGCN~\cite{cheng2023multi}, and PKEF~\cite{meng2023parallel} focus on integrating heterogeneous behavioral signals through graph propagation or multi-task learning. In addition, we include several recent \textbf{\textit{robustness-oriented approaches}} that combine contrastive or invariant learning principles, such as S-MBRec~\cite{gu2022self}, MBSSL~\cite{xu2023multi}, MISSL~\cite{wu2024multi}, and UIPL~\cite{yan2025user}, which represent the latest trend toward improving stability under noisy or inconsistent supervision. These baselines cover both traditional fusion paradigms and recent robustness-driven designs, providing a comprehensive benchmark to evaluate the effectiveness of~\shortname. 

For evaluation, we adopt the standard leave-one-out strategy~\cite{jin2020multi}, reserving the latest interaction of each user for testing and one additional instance for validation. 
Full-ranking evaluation~\cite{He2020LightGCNSA} is employed to ensure fairness, and performance is reported using HR@K and NDCG@K~\cite{herlocker2004evaluating}. 
All models are implemented in PyTorch~\cite{paszke2019pytorch} with the Adam optimizer~\cite{kingma2014adam}, and results are averaged over five random seeds for reliability.
Detailed dataset descriptions,  baseline descriptions, hyperparameter settings, and implementation details are provided in Appendix~\ref{appendix:baselines}.

\subsection{Overall Performance}

\begin{table*}[t]
\vspace{-0.2cm}
\small
\centering
\caption{Overall performance comparison in terms of HR@K and NDCG@K, where $K=10$ and $K=20$. The best results are highlighted in \textbf{bold}, and the second-best are \underline{underlined}. Results marked with * indicate statistical significance with $p$-value $<$ 0.05 over the best baseline (paired t-test).}
\label{tab:overall}
\resizebox{\linewidth}{!}{
\begin{tabular}{l|l|cc|cccccccc|cc}
\toprule
\multirow{2}{*}{\textbf{Dataset}} & \multirow{2}{*}{\textbf{Metric}} 
& \multicolumn{2}{c|}{\textbf{Single-Behavior Models}} 
& \multicolumn{8}{c|}{\textbf{Multi-Behavior Models}} 
& \multicolumn{2}{c}{\textbf{Ours}} \\
\cmidrule(r){3-14} 
& & \textbf{LightGCN} & \textbf{SimGCL} 
  & \textbf{CRGCN} & \textbf{MB-CGCN} & \textbf{PKEF} & \textbf{BCIPM} & \textbf{MISSL} & \textbf{S-MBRec} & \textbf{MBSSL} & \textbf{UIPL} & \textbf{\shortname} &  \textbf{\textit{Imp.}}\\
\midrule
\multirow{4}{*}{\textbf{Taobao}}  
 & \textbf{HR@10}   &0.0253 &0.0400 &0.1142 &0.0989 &0.1098 &\underline{0.1257} &  0.0782&0.0568 &0.0861 &0.1225 &\textbf{0.1462*} &\textbf{16.3\%} \\
 & \textbf{NDCG@10} &0.0137 &0.0236 &0.0632 &0.0471 &0.0627 &0.0708 & 0.0629 &0.0328 &0.0480 &\underline{0.0748} &\textbf{0.0969*} &\textbf{29.5\%} \\
 & \textbf{HR@20}   &0.0364 &0.0522 &0.1477 &0.1368 &0.1223 &\underline{0.1723} & 0.1134 &0.0891 &0.1247 &0.1567 &\textbf{0.1766*} &\textbf{2.5\%} \\
 & \textbf{NDCG@20} &0.0178 &0.0265 &0.0773 &0.0595 &0.0651 &0.0833 & 0.0759 &0.0411 &0.0579 &\underline{0.0834} &\textbf{0.1046*} &\textbf{25.4\%} \\
\midrule
\multirow{4}{*}{\textbf{TMall}}  
 & \textbf{HR@10}   &0.0379 &0.0709 &0.0840 &0.1083 &0.1138 &0.1418 & 0.0691 &0.0694 &0.0760 &\underline{0.1484} &\textbf{0.1549*} &\textbf{4.3\%} \\
 & \textbf{NDCG@10} &0.0205 &0.0399 &0.0442 &0.0426 &0.0642 &0.0743 & 0.0552 &0.0362 &0.0421 &\underline{0.0819} &\textbf{0.0838*} &\textbf{2.3\%} \\
 & \textbf{HR@20}   &0.0486 &0.1018 &0.1238 &0.1370 &0.1676 &\underline{0.2025} & 0.1012 &0.1009 &0.1113 &0.2005 &\textbf{0.2132*} &\textbf{5.2\%} \\
 & \textbf{NDCG@20} &0.0231 &0.0487 &0.0540 &0.0523 &0.0562 &0.0891 &0.0667  &0.0438 &0.0509 &\underline{0.0946} &\textbf{0.0982*} &\textbf{3.8\%} \\
\midrule
\multirow{4}{*}{\textbf{Beibei}}  
 & \textbf{HR@10}   &0.0383 &0.0420 &0.0539 &0.0579 &\underline{0.0813} &0.0670 & 0.0569 &0.0489 &0.0626 &0.0731 &\textbf{0.0924*} &\textbf{13.6\%} \\
 & \textbf{NDCG@10} &0.0207 &0.0210 &0.0259 &0.0381 &\underline{0.0421} &0.0331 &0.0391 &0.0253 &0.0352 &0.0377 &\textbf{0.0489*} &\textbf{16.1\%} \\
 & \textbf{HR@20}   &0.0607 &0.0665 &0.0944 &0.1073 &\underline{0.1213} &0.1014 & 0.0850 &0.0770 &0.0935 &0.1131 &\textbf{0.1362*} &\textbf{12.2\%} \\
 & \textbf{NDCG@20} &0.0262 &0.0273 &0.0361 &0.0428 &\underline{0.0525} &0.0451 &0.0499 &0.0324 &0.0419 &0.0482 &\textbf{0.0599*} &\textbf{14.1\%} \\
\bottomrule
\end{tabular}}
\end{table*}

\begin{table*}[t]
\vspace{-0.2cm}
\centering
\caption{Ablation study of ~\shortname ~ on three datasets in terms of HR@10 and NDCG@10. Values in parentheses indicate relative drops compared to the full model.}
\label{tab:ablation}
\begin{tabular}{l|cc|cc|cc}
\toprule
\multirow{2}{*}{\textbf{Model Variant}} & \multicolumn{2}{c|}{\textbf{Taobao}} & \multicolumn{2}{c|}{\textbf{TMall}} & \multicolumn{2}{c}{\textbf{Beibei}} \\
\cmidrule(lr){2-3} \cmidrule(lr){4-5} \cmidrule(lr){6-7}
 & \textbf{HR@10} & \textbf{NDCG@10 }& \textbf{HR@10} &\textbf{ NDCG@10} & \textbf{HR@10 }& \textbf{NDCG@10} \\
\midrule
\textbf{\shortname} & \textbf{0.1462*} & \textbf{0.0969*}& 0.1549* & 0.0838*  & \textbf{0.0924*} & \textbf{0.0489*} \\
\midrule
\textbf{w/o RRM}   &0.1089 { (-25.5\%)} &0.0702 { (-27.5\%)} &0.1083 { (-30.1\%)} &0.0589 { (-29.7\%)} &0.0667 { (-27.8\%)} &0.0321 { (-34.3\%)} \\
\textbf{w/o ORM}  &0.1321 { (-9.6\%)}  &0.0887 { (-8.4\%)}  &0.1360 { (-12.2\%)} &0.0766 { (-8.6\%)}  &0.0871 { (-5.7\%)}  &0.0461 { (-5.7\%)} \\
\textbf{w/o View} &0.1093 { (-25.2\%)} &0.0758 { (-21.7\%)} &0.0686 { (-55.7\%)} &0.0411 { (-51.0\%)} &0.0789 { (-14.6\%)} &0.0425 { (-12.9\%)} \\
\textbf{w/o Add-to-cart} &0.1016 { (-30.5\%)} &0.0603 { (-37.7\%)} &\textbf{0.1895 { (+22.3\%)}} &\textbf{0.1043 { (+24.5\%)}} &0.0571 { (-38.1\%)} &0.0294 { (-39.7\%)} \\
\bottomrule
\end{tabular}
\vspace{-0.2cm}
\end{table*}

We present the overall comparison results on three benchmark datasets in Table~\ref{tab:overall}. 
Across almost all metrics,~\shortname~consistently surpasses both single- and multi-behavior baselines, indicating its strong ability to utilize auxiliary signals while remaining resilient to behavioral noise and heterogeneity.

\begin{itemize}[leftmargin=0.3cm, itemindent=0cm]

    \item  \textbf{Multi-behavior modeling consistently outperforms single-behavior methods.} LightGCN and SimGCL, which rely solely on the target behavior, achieve noticeably lower accuracy compared to multi-behavior models. This validates the effectiveness of incorporating auxiliary behavioral signals, which provide complementary supervision and alleviate target sparsity. 

    \item  \textbf{\shortname~achieves the best overall results and maintains robustness across datasets.} It consistently surpasses all baselines, including the recent robustness-oriented UIPL, whose environment-level invariance constraint yields the second-best performance. The comparison suggests that explicitly modeling each behavior as an independent environment—rather than constructing synthetic ones as in UIPL—enables~\shortname~to better capture genuine behavioral heterogeneity. The local alignment (RRM) and global invariance (ORM) modules work collaboratively to ensure stable optimization and effective information transfer, leading to both higher accuracy and robustness. 

    \item \textbf{Stable and consistent performance across diverse datasets.} The performance trends of~\shortname~remain stable under different behavioral characteristics. On Taobao, where auxiliary behaviors are moderately noisy,~\shortname~achieves substantial gains ( +29.5\% on NDCG@10). On TMall, where behaviors are weakly correlated with the target, it maintains competitive performance without degradation. Even on the clean and well-structured Beibei, it continues to outperform all baselines. These results highlight the strong generalization ability of ~\shortname~and its robustness to varying behavior quality and distributional patterns. 

\end{itemize}

\subsection{Ablation Study}

To better understand the contribution of each component in~\shortname, we conduct ablation experiments at both the module and behavior levels. 
At the module level, we remove the \textbf{Representation Robustness Module (RRM)} and the \textbf{Optimization Robustness Module (ORM)} to isolate their respective effects on local and global robustness. 
At the behavior level, we remove individual auxiliary behaviors (view and add-to-cart) to examine how different interaction types contribute to target behavior prediction. 
The results are summarized in Table~\ref{tab:ablation}.

\begin{itemize}[leftmargin=0.3cm, itemindent=0cm]

    \item \textbf{Both RRM and ORM are essential for robustness, jointly enhancing performance through local–global collaboratively.} Removing either module leads to consistent performance degradation across all datasets. Eliminating RRM causes a sharp decline (e.g., Taobao: $-25.5\%$ HR@10, $-27.5\%$ NDCG@10), verifying that local semantic consistency via contrastive learning is crucial for mitigating semantic drift in auxiliary behaviors. Similarly, removing ORM also degrades results (e.g., TMall: $-12.2\%$ HR@10), showing that global invariance regularization effectively stabilizes learning across heterogeneous behavioral environments. These results confirm that RRM and ORM work in a collaborative manner: RRM focuses on local semantic consistency, while ORM provides global optimization stability, and their joint optimization yields the most robust performance.
    
    \item \textbf{Auxiliary behaviors substantially enhance recommendation performance in most cases.} Removing auxiliary interactions generally leads to notable drops in both HR@10 and NDCG@10, indicating that multi-behavior information provides valuable complementary supervision. For example, removing view interactions on Taobao reduces HR@10 by $25.2\%$, and removing add-to-cart on Beibei causes a $38.1\%$ decline. These results demonstrate that auxiliary signals can significantly enrich user representation learning and that ~\shortname~can effectively extract useful information from them through target-anchored alignment and invariance optimization. 

    \item \textbf{Performance degradation from noisy auxiliary behaviors.} We observe that incorporating certain auxiliary behaviors can, counter-intuitively, harm performance. A notable example is on the TMall dataset, where removing the \textit{add-to-cart} behavior results in a significant performance gain (HR@10 +22.3\%). This phenomenon is consistent with its extremely low Behavioral Alignment Ratio (BAR = 0.0012), indicating a severe semantic misalignment with the target purchase behavior. In such cases, the auxiliary signal acts primarily as noise, leading to negative transfer. The fact that ~\shortname~maintains robust performance even in the presence of such misleading signals underscores the effectiveness of its RRM-ORM collaborative design. This finding critically highlights the necessity for future designs to move beyond uniform behavior integration towards more adaptive or selective mechanisms.
    
\end{itemize}

We further investigate the impact of different invariance regularization strategies for the ORM module. Empirical results (detailed in Section~\ref{sec:invariance_analysis}) demonstrate that \textbf{Risk Extrapolation (REx)}~\cite{krueger2021out} consistently outperforms gradient-based IRM variants (IRM-V1 and IRM-V2) by enforcing risk consistency across heterogeneous behaviors. Consequently, we adopt REx to ensure global optimization stability in ~\shortname.

\subsection{Robustness Study}

To further examine the robustness of~\shortname~under noisy environments, we conduct a controlled perturbation study on the Taobao dataset by randomly modifying auxiliary behavior interactions. 
Specifically, we randomly \textbf{add 10\%, 30\%, and 50\% new edges} to simulate spurious behaviors and \textbf{remove 10\%, 30\%, and 50\% existing edges} to simulate missing auxiliary information, enabling evaluation under both over- and under-observed conditions. 
We compare~\shortname~with three competitive multi-behavior baselines—UIPL~\cite{yan2025user}, BCIPM~\cite{yan2024behavior}, and PKEF~\cite{meng2023parallel}—that perform strongly under standard settings.

\begin{figure}
    \centering
    \includegraphics[width =\linewidth]{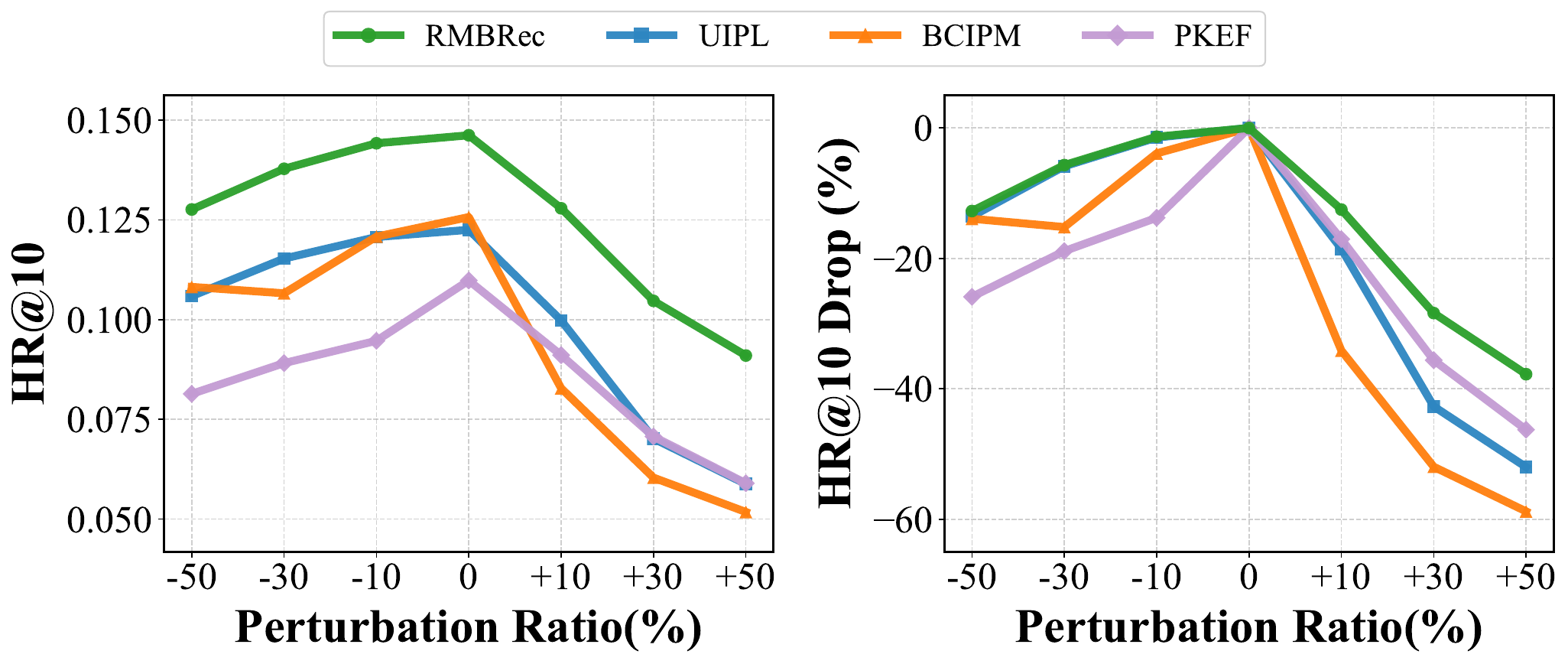}
    \caption{{Robustness analysis under different perturbation ratios on the Taobao dataset (HR@10).}}
    \label{fig:robustness}
    \vspace{-0.2cm}
\end{figure}

As shown in Figure~\ref{fig:robustness}, the left plot reports absolute performance (HR@10), while the right plot shows the relative drop ratio compared to the noise-free condition. ~\shortname~consistently achieves the best performance across all perturbation ratios, regardless of whether noise is introduced by excessive or missing interactions, and exhibits a substantially smaller performance decline than competing methods. This robustness arises from the collaborative effect of the \textbf{Representation Robustness Module (RRM)}, which suppresses semantically drifted auxiliary behaviors, and the \textbf{Optimization Robustness Module (ORM)}, which enforces global invariance across disturbed behavioral environments, enabling stable and reliable recommendation performance under severe noise.

\subsection{Hyperparameter Sensitivities}

\begin{figure}
    \centering
    \includegraphics[width =\linewidth]{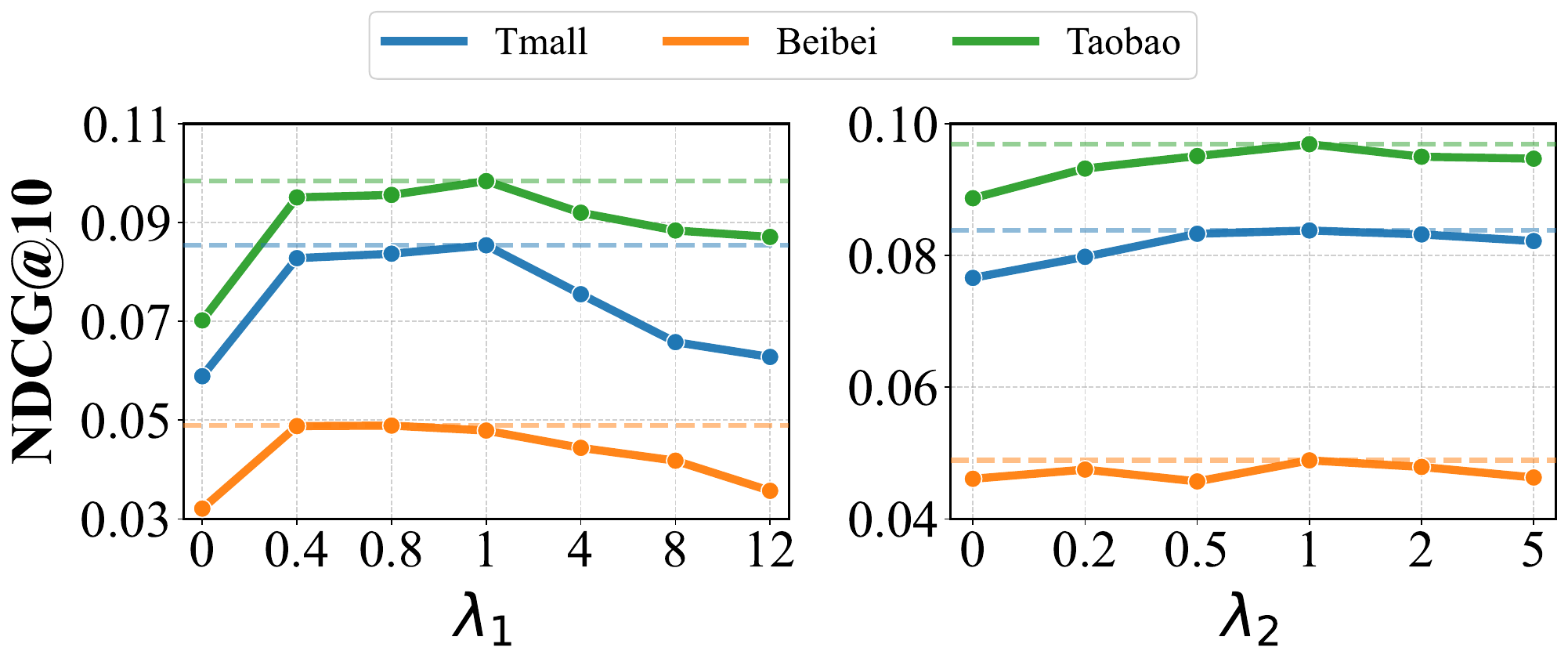}
    \caption{Hyperparameter sensitivity analysis of ~\shortname ~ with $\lambda_1$ and $\lambda_2$.}
    \label{fig:hyper}
    \vspace{-0.2cm}
\end{figure}

We study the sensitivity of the two key hyperparameters in ~\shortname~, $\lambda_{1}$ and $\lambda_{2}$, which control the weights of contrastive alignment and invariance regularization. They jointly determine the balance between local semantic consistency and global optimization stability. The results are shown in Figure~\ref{fig:hyper}.

Across all datasets, performance first increases as $\lambda_{1}$ and $\lambda_{2}$ grow from small values, indicating that both modules effectively enhance robustness by aligning auxiliary behaviors locally and stabilizing optimization globally. This verifies the complementary roles of RRM and ORM in improving resistance to noisy or misaligned supervision. However, when either weight becomes overly large, the auxiliary objectives dominate the main recommendation loss, leading to performance degradation. These findings suggest that robustness in multi-behavior recommendation relies not on maximizing auxiliary regularization, but on maintaining a balanced integration between local alignment and global invariance.

\section{Conclusion and Future Work}

In this paper, we tackled the challenge of behavioral misalignment and noise in multi-behavior recommendation. Unlike prior fusion-based methods, we proposed~\shortname, a target-centric framework that enhances robustness through a local–global collaborative mechanism: the Representation Robustness Module (RRM) enforces local semantic consistency via target-anchored contrastive alignment, while the Optimization Robustness Module (ORM) ensures global optimization stability by learning invariant user preferences across behavioral environments. Extensive experiments on three real-world datasets verified the superiority of~\shortname~in both accuracy and robustness. Moreover, the proposed Behavioral Alignment Ratio (BAR) offers a simple yet effective diagnostic metric for quantifying behavioral consistency and guiding robustness-oriented model design. 

Although~\shortname~shows strong robustness, our analysis also reveals open challenges. On TMall, the noisy add-to-cart behavior still degrades performance, suggesting that alignment may be suboptimal under weak behavioral correlations. Future work will explore adaptive alignment guided by data characteristics (e.g., BAR) to selectively emphasize informative signals. We also plan to extend the proposed local–global robustness framework to model temporal dynamics and broader multi-source recommendation scenarios. 

\begin{acks}
This research is supported by the New Cornerstone Science Foundation through the XPLORER PRIZE and the National Natural Science Foundation of China (No. 62272254, No. U25A20445, No. 72188101).
\end{acks}

\newpage

\bibliographystyle{ACM-Reference-Format}
\bibliography{sample-base}

\appendix

\section{Further Model Analysis and Discussion}
\label{appendix:model}

\subsection{Information-Theoretic Analysis}
\label{appendix:it_view}
Let $P_u^{(b)}$ denote the user embedding learned under behavior $b\in\mathcal{B}$, and $P_u^{(t)}$ the user embedding under the target behavior.
Let $Y_t$ denote the target label.
From an information-theoretic viewpoint, robust multi-behavior learning should (i) \textbf{maximize the shared information between auxiliary and target user representations (local purification)}, and (ii) \textbf{stabilize the target-relevant information across heterogeneous behavioral environments (global invariance)}.

\textbf{RRM as local MI alignment.}
RRM enforces a target-anchored alignment by contrasting $(P_u^{(b)}, P_u^{(t)})$ with in-batch negatives. Concretely, the InfoNCE-style objective used in Eq.(3) is a lower bound estimator of the mutual information $I(P_u^{(b)}; P_u^{(t)})$. Therefore, increasing the contrastive score improves a valid lower bound of this MI and reduces the conditional uncertainty $H(P_u^{(b)}\!\mid P_u^{(t)})$. Formally, letting $\text{sim}(\cdot,\cdot)$ be cosine similarity and $\tau>0$ the temperature, the loss in Eq.(\ref{eq:rrm_loss_unit}) satisfies 
\[
\underbrace{I\!\big(P_u^{(b)}; P_u^{(t)}\big)}_{\text{shared info}}
\;\;\ge\;\;
\mathbb{E}\!\left[
\log \frac{\exp\!\big(\text{sim}(p_u^{(b)}, p_u^{(t)})/\tau\big)}
{\sum_{u'} \exp\!\big(\text{sim}(p_u^{(b)}, p_{u'}^{(b)})/\tau\big)}
\right],
\]
hence maximizing the RRM (minimizing Eq.(\ref{eq:rrm_loss_unit})) tightens a lower bound to $I(P_u^{(b)}; P_u^{(t)})$ and purifies auxiliary representations toward target semantics.

\textbf{ORM as global risk-variance invariance.}
While RRM focuses on enhancing the mutual information between auxiliary and target representations, ORM aims to ensure that the predictive information carried by user embeddings remains stable across behaviors. We treat each behavior $b\!\in\!\mathcal{B}$ as a distinct environment with its own label signal $Y^{(b)}$, representing behavior-specific supervision (e.g., click, cart, or purchase). Under this view, the empirical mutual information between the user representation and its local label is $ I^{(b)} = I(P_u^{(b)}; Y^{(b)})$, and a robust model should maintain invariant predictive information across environments: 
\[ 
\min_{\Theta}\;\mathrm{Var}_{b\in\mathcal{B}}\!\big[I(P_u^{(b)}; Y^{(b)})\big]. 
\] 
Since directly estimating $I(P_u^{(b)}; Y^{(b)})$ is intractable, we adopt an optimization surrogate: the empirical variance of per-behavior BPR risks. As lower risk variance indicates more consistent predictive information among behaviors, the ORM loss (Eq.~(\ref{eq:orm_loss})) is:
\[
\mathcal{L}_{\mathrm{ORM}}
=\mathrm{Var}\!\big(\{\mathcal{L}^{(b)}_{\mathrm{BPR}}\}_{b\in\mathcal{B}}\big)
=\frac{1}{|\mathcal{B}|-1}\!\sum_{b\in\mathcal{B}/target}\!(\mathcal{L}^{(b)}_{\mathrm{BPR}}-\bar {\mathcal{L}})^2.
\]
This regularizer reduces the variation of training risks across behaviors, which effectively stabilizes the mutual information $I(P_u^{(b)};Y^{(b)})$ among behavioral environments. By constraining such variation, ORM prevents the model from overfitting to behavior-specific supervision signals and encourages representations that preserve predictive information in a consistent, invariant manner. 

\textbf{Collaborative robustness objective.} Combining the above two principles, the overall information-theoretic objective of RMBRec can be expressed as: 

\[
\max_{\Theta}\; \Big( \underbrace{\mathbb{E}_{b\in\mathcal{B}_{\text{aux}}} [I(P_u^{(b)};P_u^{(t)})]}_{\text{(1) local MI alignment}} - \lambda\,\underbrace{\mathrm{Var}_{b\in\mathcal{B}} [I(P_u^{(b)};Y^{(b)})]}_{\text{(2) global MI invariance}} \Big). 
\] 

This unified formulation connects the representation and optimization level perspectives through mutual information. In particular, the first term encourages auxiliary embeddings to share consistent semantics with the target behavior (\textit{local semantic consistency}), while the second term enforces cross-behavior stability of the predictive information (\textit{global optimization stability}).

\subsection{Detailed Description of ~\shortname ~Algorithm}
\label{appendix:algorithm}

\begin{algorithm}[t]
\renewcommand{\algorithmicrequire}{\textbf{Input:}}
\renewcommand{\algorithmicensure}{\textbf{Output:}}
\caption{\small{The Algorithm of ~\shortname}.}\label{alg:rmbrec}
\begin{algorithmic}[1]
\REQUIRE Multi-behavior interactions $\{\textbf{R}^{(b)}\}_{b\in\mathcal{B}}$, learning rate $\eta$, hyperparameters $\lambda_1$, $\lambda_2$;  ~~\\
\ENSURE Trained model parameters $\Theta$; ~~\\
\STATE Initialize final user embedding matrix $\mathbf{Z}_u \in \mathbb{R}^{|\mathcal{U}| \times d}$; \\
\STATE Initialize final item embedding matrix $\mathbf{Z}_i \in \mathbb{R}^{|\mathcal{I}| \times d}$; \\
\FOR{$epoch=1,2,...,T$} 
    \FOR{batch in user set $\mathcal{U}$} 
    \STATE \textbf{Step 1: Learn Behavior-Specific Embeddings}
    \FOR{each behavior $b \in \mathcal{B}$}
        \STATE $\mathbf{P}^{(b)},\mathbf{Q}^{(b)},\mathcal{L}^{(b)}_{BPR} \leftarrow \text{LightGCN}(\mathbf{R}^{(b)}, \mathbf{Z}_u, \mathbf{Z}_i)$;
    \ENDFOR
    
    \STATE \textbf{Step 2: Representation Robustness Module}
    \STATE Calculate representation robustness loss $\mathcal{L}_{RRM}$ by Eqn.(\ref{eq:rrm_loss});
    
    \STATE \textbf{Step 3: Optimization Robustness Module}
    \STATE Calculate optimization robustness loss $\mathcal{L}_{RRM}$ by Eqn.~(\ref{eq:orm_loss});
    
    \STATE \textbf{Step 4: Robustness-Enhanced Fusion}
    \STATE $\mathbf{z}_u \leftarrow \frac{1}{|\mathcal{B}|} \sum_{b \in \mathcal{B}} \mathbf{p}_u^{(b)}$ for each user in batch;
    \STATE $\mathbf{z}_i \leftarrow \frac{1}{|\mathcal{B}|} \sum_{b \in \mathcal{B}} \mathbf{q}_i^{(b)}$ for each item in batch;
    
    \STATE \textbf{Step 5: Target Behavior Prediction}
    \STATE Calculate main recommendation loss $\mathcal{L}_{main}$ by Eqn.~(\ref{equa:bpr_main});
    
    \STATE \textbf{Step 6: Model Optimization}
    \STATE Calculate total loss $\mathcal{L} \leftarrow \mathcal{L}_{main} + \lambda_1 \mathcal{L}_{RRM} + \lambda_2 \mathcal{L}_{ORM}$;
    \STATE Update $\Theta \gets \Theta - \eta \cdot \nabla_{\Theta}\mathcal{L}$;
    
    \IF{early stopping condition is met}
        \STATE break;
    \ENDIF
    \ENDFOR
\ENDFOR
\STATE \textbf{return} trained parameters $\Theta$.
\end{algorithmic}
\end{algorithm}

Algorithm~\ref{alg:rmbrec} outlines the complete training procedure of ~\shortname, which systematically integrates multi-behavior learning with dual-level robustness enhancement. 

Complexity and Efficiency Analysis. We demonstrate that RMBRec is efficient and scalable. Theoretically, the time complexity is dominated by the backbone encoder ($\mathcal{O}(|\mathcal{R}|Ld)$), which scales linearly with data size. The proposed robustness modules add minimal overhead: RRM introduces a constant cost per batch ($\mathcal{O}(|\mathcal{B}|B^2 d)$), and ORM is computationally negligible. Empirically, on the Taobao dataset, RMBRec requires 16.7s per epoch. While this is naturally higher than the lightweight LightGCN (4.9s), it is approximately 3$\times$ faster than the state-of-the-art robust baseline UIPL (51.6s). This confirms that RMBRec achieves superior performance with high training efficiency.








\section{Detailed Experimental Settings}
\label{appendix:baselines}
\subsection{Datasets}

\begin{table}[t]
\centering
\caption{Statistics of the datasets used in our experiments.}
\label{appendix_tab_dataset}
\resizebox{\linewidth}{!}{
\begin{tabular}{lcccccc}
\toprule
\textbf{Dataset} & \textbf{\#Users} & \textbf{\#Items} & \textbf{\#View} & \textbf{\#Collect} & \textbf{\#Add-to-cart} & \textbf{\#Purchase} \\
\midrule
Taobao  & 48,749 & 39,493 & 1,548,162 & -        & 193,747  & 259,771 \\
TMall   & 41,738 & 11,953 & 1,813,498 & 221,514  & 1,996    & 287,158 \\
Beibei  & 21,716 & 7,977  & 2,412,586 & -       & 642,622  & 304,576 \\
\bottomrule
\end{tabular}}
\vspace{-0.5cm}
\end{table}

We provide detailed descriptions of the three datasets used in our experiments. Following established practices in prior research~\cite{yan2025user,yan2024behavior}, we retain only the earliest occurrence of each user–item interaction to remove redundant records. The detailed statistics after preprocessing are summarized in Table~\ref{appendix_tab_dataset}. 

\begin{itemize}[leftmargin=0.3cm, itemindent=0cm]

    \item \textbf{Taobao: } This dataset is obtained from a widely used Chinese e-commerce platform. It records multiple types of user behaviors, including view, add-to-cart, and purchase. Taobao exhibits medium behavioral consistency, sitting between Beibei (highly aligned) and TMall (highly noisy), and provides a balanced benchmark for evaluating generalization under diverse behavioral distributions.
    
    \item \textbf{TMall:} This dataset is collected from Tmall, one of the largest online marketplaces in China. It covers four types of behaviors, including view, collect, add-to-cart, and purchase. Unlike Beibei, TMall contains highly imbalanced and noisy auxiliary signals. For example, only a tiny fraction of add-to-cart interactions lead to purchases (BAR = 0.0012), making it ideal for evaluating robustness against noise and misalignment.
    
    \item \textbf{Beibei:}This dataset is collected from Beibei, an e-commerce platform specializing in maternal and infant products. It contains three types of interactions, including view, add-to-cart, and purchase. Compared to TMall and Taobao, Beibei shows highly consistent user intentions—most purchases are preceded by auxiliary actions, as confirmed by its high BAR values (view = 0.9756, cart = 0.9731).
    
\end{itemize}

\subsection{Baselines}
\label{appendix:baselines}
We summarize below the detailed introductions of all baselines compared in our experiments. They can be classified into two main categories: \textbf{single-behavior methods}, which only model the target behavior (e.g., purchases), and \textbf{multi-behavior methods}, which explicitly leverage auxiliary interactions. Among them, several recent approaches (SimGCL, MBSSL, UIPL) also aim to improve model robustness against noisy supervision or distributional shifts. 

\noindent\textbf{Single-behavior Models.}
\begin{itemize}[leftmargin=0.3cm, itemindent=0cm]

    \item \textbf{LightGCN}~\cite{He2020LightGCNSA}: A simplified graph-based collaborative filtering model that removes feature transformation and non-linear activations, effectively capturing high-order user–item relations on the interaction graph.
    
    \item \textbf{SimGCL}~\cite{Yu2021AreGA}: It introduces random embedding perturbations and an InfoNCE-based regularization. This enhances representation robustness and generalization, making it a strong baseline for robustness comparison.
    
\end{itemize}

\noindent\textbf{Multi-behavior Models.}

\begin{itemize}[leftmargin=0.3cm, itemindent=0cm]

    
    \item \textbf{CRGCN}~\cite{yan2023cascading}: It propagates embeddings across behaviors in sequence, refining user preferences through multi-task learning that jointly optimizes auxiliary and target tasks.
    
    \item \textbf{MB-CGCN}~\cite{cheng2023multi}: It simplifies residual propagation and aggregates embeddings across all behaviors to strengthen target recommendation.
    
    \item \textbf{PKEF}~\cite{meng2023parallel}: It is a parallel knowledge fusion framework that employs a disentangled multi-expert structure to balance heterogeneous behavior distributions and reduce negative transfer effects.
    
    \item \textbf{BCIPM}~\cite{yan2024behavior}: It is a behavior-contextualized item preference model that leverages auxiliary interactions to enhance target recommendation while filtering out noise from irrelevant behaviors with limited invariance guarantees.

    \item \textbf{MISSL}~\cite{wu2024multi}: It is a unified multi-behavior and multi-interest framework that employs a hypergraph transformer to extract behavior-specific and shared interests, enhanced by self-supervised contrastive learning and a behavior-aware training task for stable optimization.
    
    \item \textbf{S-MBRec}~\cite{gu2022self}: It alleviates target sparsity by introducing auxiliary contrastive objectives. It adopts a star-shaped design that provides additional training signals without explicitly modeling inter-auxiliary dependencies.
    
    \item \textbf{MBSSL}~\cite{xu2023multi}: It employs both inter- and intra-behavior contrastive learning to enhance semantic consistency. It further adopts adaptive gradient balancing to maintain stable optimization, thus improving robustness.
    
    \item \textbf{UIPL}~\cite{yan2025user}: It is the first attempt to incorporate invariant risk minimization into multi-behavior recommendation. It applies a VAE-based invariance constraint. However, its synthetic environment design limits its ability to capture realistic behavioral heterogeneity under distribution shifts.

\end{itemize}

\section{More Experimental Analysis}
\label{appendix:exp_analysis}

\subsection{Further Robustness Analyses}

\label{appd:robustness}

\begin{figure}
    \centering
    \includegraphics[width =\linewidth]{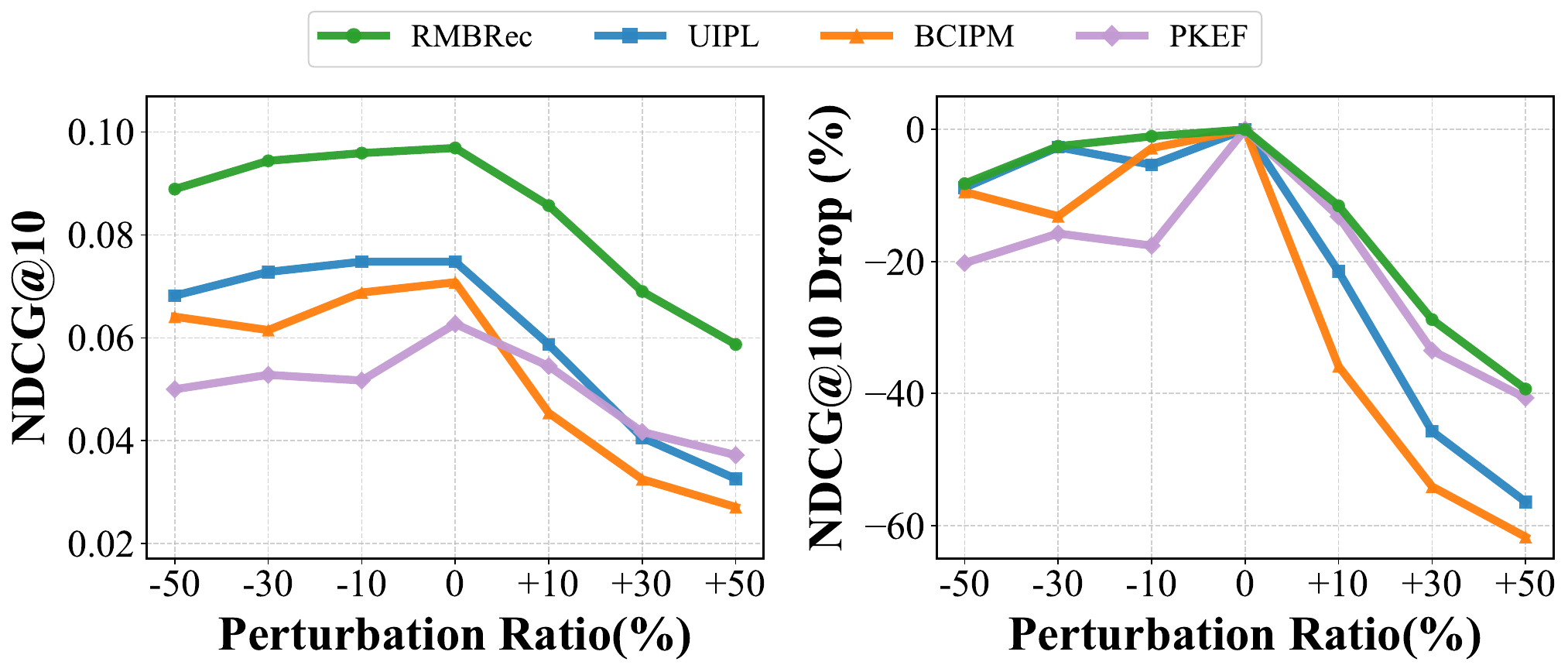}
    \caption{{Robustness analysis under different perturbation ratios on the Taobao dataset (NDCG@10).}}
    \label{fig:robustness3}
    \vspace{-0.5cm}
\end{figure}

To complement the robustness analysis reported in the main text (Figure~\ref{fig:robustness}), which focuses on \textbf{HR@10}, we additionally report the results in terms of \textbf{NDCG@10} in Figure~\ref{fig:robustness3}. The overall trend is consistent with the HR-based findings: as the perturbation ratio increases, all models experience performance degradation, yet~\shortname~maintains the highest ranking quality and the smallest drop across all noise levels. This further confirms that the proposed \textit{Representation Robustness Module (RRM)} and \textit{Optimization Robustness Module (ORM)} jointly improve stability by filtering semantically drifted auxiliary behaviors and enforcing invariance across disturbed environments. The alignment between HR and NDCG results demonstrates that~\shortname~achieves not only higher hit accuracy but also more reliable ranking robustness under noisy conditions across behavior types.

\subsection{Detailed Description of IRM Variants}

\label{app:irm}

\subsubsection{Background on IRM}
Invariant Risk Minimization (IRM)~\cite{arjovsky2019invariant} was proposed as a principle for out-of-distribution (OOD) generalization~\cite{liu2021towards,wang2022causal}. Its key idea is that if a representation truly captures invariant and causal factors, then a single linear classifier should perform well across multiple environments~\cite{bai2024multimodality,chen20253}. Formally, IRM aims to find a feature extractor $\Phi$ and a classifier $w$ such that~\cite{arjovsky2019invariant}: 
\begin{equation} 
    \min_{\Phi, w} \; \sum_{e \in \mathcal{E}} R^e(w \circ \Phi), \quad \text{s.t. } w \in \arg\min_{\tilde{w}} R^e(\tilde{w} \circ \Phi), \ \forall e, 
\end{equation} 
where $R^e(\cdot)$ denotes the risk in environment $e$. This constraint enforces that the same classifier is simultaneously optimal across environments, encouraging $\Phi$ to discard spurious correlations and preserve invariant features~\cite{ahuja2020invariant}. Since solving this constrained problem directly is intractable, several relaxations have been developed~\cite{arjovsky2019invariant}. We summarize three widely used ones: IRMv1~\cite{arjovsky2019invariant}, IRMv2~\cite{ahuja2020invariant}, and Risk Extrapolation (REx)~\cite{krueger2021out}. 

\subsubsection{IRMv1}

IRMv1 is the most widely adopted implementation of invariant risk minimization~\cite{arjovsky2019invariant}. Instead of strictly enforcing the ideal constraint, it introduces a penalty that aligns the gradients of the risks with respect to the classifier across environments:
\begin{equation} 
    \min_{\Phi, w} \; \sum_{e \in \mathcal{E}} R^e(w \circ \Phi) + \lambda \cdot \sum_{e \in \mathcal{E}} \big\| \nabla_{w|w=1} R^e(w \circ \Phi) \big\|^2, 
\end{equation}
where $\lambda$ controls the strength of the penalty. This formulation encourages $\Phi$ to learn features that support a common classifier shared across environments. Although principled, the gradient-based penalty is known to be difficult to optimize, and its effectiveness may degrade in noisy or high-variance settings.

\subsubsection{IRMv2}

IRMv2~\cite{ahuja2020invariant} offers a simplified implementation of invariant risk minimization by fixing the classifier parameter (e.g., $w=1$) and penalizing the gradients of the risk under this fixed setting for efficiency and robustness: 
\begin{equation} 
    \min_{\Phi} \; \sum_{e \in \mathcal{E}} R^e(w \circ \Phi) + \lambda \sum_{e \in \mathcal{E}} \Big\| \nabla_{w \mid w=1} R^e(w \circ \Phi) \Big\|^2 .
\end{equation}
This relaxation avoids the costly bi-level optimization in the original IRM objective, making it easier to implement and computationally more stable~\cite{ahuja2020invariant,liao2025mitigating,yang2025invariance}. However, since the classifier is no longer learnable, the model may lose adaptability, and the invariance constraint is only approximately enforced, which often results in weaker performance compared with IRMv1. 

\subsubsection{REx}

Risk Extrapolation (REx)~\cite{krueger2021out} departs from gradient alignment and instead enforces invariance by directly minimizing the variance of risks across environments:
\begin{equation} 
    \min_{\Phi, w} \; \frac{1}{|\mathcal{E}|} \sum_{e \in \mathcal{E}} R^e(w \circ \Phi) + \lambda \cdot \mathrm{Var}_{e \in \mathcal{E}} \big(R^e(w \circ \Phi)\big). 
\end{equation} 
This design encourages consistent performance across environments and avoids the optimization difficulties of IRMv1. Empirically, it tends to yield more robust and stable results, though it enforces invariance only at the risk level rather than directly constraining representations or gradients.

In summary, IRMv1 provides a principled approximation of the ideal IRM objective but suffers from optimization challenges. IRMv2 improves simplicity and efficiency at the expense of fidelity to the original principle. REx offers a more stable and practical alternative by encouraging risk-level consistency, which often translates into stronger robustness in practice. In our framework, we examine all three variants to better understand their relative strengths for multi-behavior recommendation. 

\subsection{Impact of Invariance Regularization Variants}
\label{sec:invariance_analysis}

To identify the most effective implementation for our Optimization Robustness Module (ORM), we examine three widely used variants of Invariant Risk Minimization (IRM)~\cite{arjovsky2019invariant}: IRM-V1~\cite{arjovsky2019invariant}, IRM-V2~\cite{ahuja2020invariant}, and IRM-REx~\cite{krueger2021out}. The comparison results are reported in Table~\ref{tab:inv}.

\begin{table}[t]
    \centering
    \caption{Comparison of different invariance regularization variants in \shortname. }
    \label{tab:inv}
    \resizebox{\linewidth}{!}{
        \begin{tabular}{l|cc|cc|cc}
        \toprule
        \multirow{2}{*}{\textbf{Variant}} & \multicolumn{2}{c|}{\textbf{Taobao}} & \multicolumn{2}{c|}{\textbf{TMall}} & \multicolumn{2}{c}{\textbf{Beibei}} \\
        \cmidrule(lr){2-3} \cmidrule(lr){4-5} \cmidrule(lr){6-7}
         & \textbf{HR@10} & \textbf{NDCG@10 }& \textbf{HR@10} &\textbf{ NDCG@10} & \textbf{HR@10 }& \textbf{NDCG@10} \\
        \midrule
        \textbf{w/o ORM}   &0.1321  &0.0887  &0.1360  &0.0766  &0.0871  &0.0461  \\
        \midrule
        \textbf{IRM-V1}  &0.1456  &0.0962  &0.1411  &0.0794  &0.0913  &0.0483  \\
        \textbf{IRM-V2 } &0.1389  &0.0938  &0.1304  &0.0729  &0.0902  &0.0478  \\
        \textbf{IRM-REx} &\textbf{0.1462}  &\textbf{0.0969}  &\textbf{0.1549}  &\textbf{0.0838}  &\textbf{0.0924}  &\textbf{0.0489}  \\
        \bottomrule
        \end{tabular}
    }
    \vspace{-0.5cm}
\end{table}

As shown in Table~\ref{tab:inv}, all IRM variants substantially outperform the base model without ORM, verifying that invariance regularization effectively suppresses behavior-specific noise. Among them, \textbf{IRM-REx} achieves the best overall performance. This advantage stems from its variance-based penalty, which enforces consistent risks across behavioral environments more stably than the gradient-based penalties in IRM-V1. In contrast, IRM-V2 performs the weakest due to its fixed classifier ($w=1$), which limits the flexibility of invariant representation learning. Based on these findings, we adopt REx as the core instantiation of ORM in ~\shortname.

\end{document}